\pdfoutput=1
\PassOptionsToPackage{dvipsnames}{xcolor}
\documentclass[letterpaper,11pt]{article}

\usepackage{jcappub}
\usepackage[T1]{fontenc}

\makeatletter
\gdef\@fpheader{}
\makeatother

\usepackage{natbib}
\usepackage{cancel}
\usepackage{enumitem}
\usepackage{amsthm}
\usepackage{amsmath} 	
\usepackage{amssymb}
\usepackage{amsfonts}
\usepackage{graphicx}
\usepackage{dblfloatfix}	
\usepackage{verbatim}
\usepackage{epstopdf}
\usepackage{tensor}
\usepackage{epsfig,multicol,bbm}
\usepackage{color}
\usepackage{colortbl}
\usepackage{float}
\usepackage{multirow}
\usepackage{array}
\usepackage{ulem}

\usepackage{bm}
\usepackage{afterpage}
\usepackage{enumitem}

\usepackage[margin=1in]{geometry}
\usepackage{tabularx}
\usepackage{adjustbox}
\newcolumntype{L}{>{\raggedright\arraybackslash}p{0.30\linewidth}}
\newcolumntype{Y}{>{\raggedright\arraybackslash}X}

\renewcommand{\arraystretch}{1.85}
\newcommand{\redsout}{\bgroup\markoverwith{\textcolor{red}{\rule[0.5ex]{2pt}{0.8pt}}}\ULon}

\usepackage{mathrsfs}
\usepackage{tikz}
\usepackage{subcaption}
\usetikzlibrary{arrows.meta}
\usepackage{hyperref}
\usepackage{boxedminipage}
\allowdisplaybreaks
\usepackage{booktabs}
\usepackage[dvipsnames]{xcolor}
\newcommand{\final}[1]{\textcolor{black}{#1}}

\title{Anti-Ultralocality and Plateau Models of Inflation}

\abstract{
Anti-ultralocality refers to the growth of spatial gradient terms relative to velocity terms in the coupled Einstein--scalar field equations. It is a characteristic feature of decelerated expansion before the onset of inflation. Previous numerical relativity studies have shown that anti-ultralocality prevents the onset of inflation in models with power-law inflaton potentials. In this paper, we show that models with plateau-shaped inflaton potentials, which are considered to be the simplest way to generate a tensor-to-scalar ratio below current observational upper limits, are especially vulnerable to anti-ultralocality effects. The reasons are the flatness of the plateau and the energy density gap of  $\sim 10$ orders of magnitude  between the Planck density and the plateau potential energy. To study the problem, we develop a protocol for assessing the viability of inflationary models in general, and we apply it to a plateau potential using a previously validated numerical relativity code. We find that, starting from generic initial conditions, the growth of gradient terms in the Einstein equations relative to non-gradient terms either prevents inflation from lasting for enough $e$-folds or triggers a phase of quantum runaway. We show that the fine-tuning of initial conditions necessary to avoid these issues becomes more severe as the energy scale of inflation is made smaller, disfavoring common approaches for reducing the tensor-to-scalar ratio.

}


\author[1]{Joshua Shterenberg,}
\author[2,3]{David Garfinkle,}
\author[1]{Anna~I.~Rosenzweig,}
\author[1,4,5]{David Shlivko,}
\author[1]{and Paul J. Steinhardt}
\affiliation[1]{Department of Physics, Jadwin Hall, Princeton University, Princeton, NJ 08544, USA}
\affiliation[2]{Department of Physics, Oakland University, Rochester, MI 48309, USA}
\affiliation[3]{Leinweber Center for Theoretical Physics, Randall Laboratory of Physics, University of Michigan, Ann Arbor, MI 48109, USA}
\affiliation[4]{Department of Physics, Stanford University, Stanford, CA 94305, U.S.A.}
\affiliation[5]{Kavli Institute for Particle Astrophysics \& Cosmology, SLAC National Accelerator Laboratory, Menlo Park, CA 94025, U.S.A.}
\emailAdd{jshteren@princeton.edu}
\emailAdd{garfinkl@oakland.edu}
\emailAdd{arosenzweig@nvidia.com}
\emailAdd{dshlivko@stanford.edu}
\emailAdd{steinh@princeton.edu}

\begin{document}
\maketitle

\section{Introduction}

The homogeneity, isotropy, and flatness of the observable universe are \final{special features,} not expected to emerge naturally from a hot initial state governed by quantum gravity.
Inflation \cite{Guth:1980zm,Linde:1981mu,Albrecht:1982wi}, an early phase of accelerated expansion, was \final{proposed} as a \final{dynamical} mechanism to smooth and flatten the universe. \final{The underlying idea was that, coming out of the hot big bang, a scalar field $\phi$, dubbed the inflaton, would render some regions of spacetime homogeneous, isotropic, and spatially flat, as described by the flat Friedmann-Robertson-Walker (FRW) metric.
These patches would then grow to overtake most of the volume and explain the flat and smooth large-scale universe we observe today.}
Yet numerical simulations of the simplest inflationary models with power-law inflaton potentials ($V(\phi) = \frac{1}{2}m^2 \phi^2$ or $\lambda \phi^4$\final{, where $m$ is the mass of the inflaton and $\lambda$} a dimensionless constant) have shown that these models fail to smooth and flatten the universe beginning from initial conditions that are \final{non-perturbatively far from} flat FRW spacetimes \cite{Garfinkle:2008ei,Goldwirth:1989pr,goldwirthInhomogeneousInitialConditions1991,goldwirthINITIALCONDITIONSINFLATION,goldwirthInitialConditionsNew1990,Garfinkle:2023vzf,Clough:2017efm,Ijjas:2024oqn,Aurrekoetxea:2019fhr}, as expected after the universe emerges from the big bang and quantum gravity dominated phase.

 \final{The} failure \final{of power-law models stems from} a fundamental feature of general relativity first \final{conjectured} by Belinski, Khalatnikov, and Lifshitz (BKL) \cite{Belinsky:1970ew,bergerSingularityGenericGravitational1998}. \final{In an expanding universe}, as the Einstein-scalar field equations are extrapolated back in time toward a \final{putative} cosmological singularity, BKL showed that all spatial gradient terms in the Einstein equations shrink and become negligible compared to non-gradient \final{(`velocity')} terms. \final{This result was} subsequently extended to the coupled Einstein–scalar field equations.
This effect, which also occurs in contracting spacetimes \cite{Ijjas:2024oqn,Ijjas:2021gkf}, is known as `ultralocality' because each point in spacetime ultimately evolves independently of its neighbors.

In the case of decelerated expansion, \final{such as the} phase between the big bang and the beginning of inflation, ultralocality manifests in reverse: spatial gradients grow and become dominant over \final{velocity} terms, driving the universe away from a homogeneous, flat-FRW state. This inverse effect is \final{dubbed} `anti-ultralocality' \cite{Ijjas:2024oqn}, and it is responsible for the failures of power-law inflation models identified in the numerical studies mentioned above \cite{Garfinkle:2023vzf}.

The goal of this paper is to study the effects of anti-ultralocality \final{in} models of inflation with `plateau potentials' of the form
\begin{align}
    V(\phi) &=V_0(1-e^{-\gamma\phi})^2,
    \label{potential}
\end{align}
where the inflaton $\phi$ is a canonical \final{scalar field} that is minimally coupled \final{to Einstein gravity}, $\gamma=\sqrt{2/3}M_{\rm Pl}^{-1}$, and $V_0 \approx 10^{-10}M_{\rm Pl}^4$. We define $M_{\rm Pl}\equiv 1/\sqrt{8\pi G}$ to be the reduced Planck mass. This form corresponds to the Starobinsky model \cite{starobinskyNewTypeIsotropic1980} and the Higgs inflation model \cite{Bezrukov:2007ep}. Simple power-law inflation models produce cosmic microwave background (CMB) temperature fluctuation spectra with a tensor-to-scalar ratio $r$ that is incompatible with the observational upper bound. The plateau model in Eq.~(\ref{potential}) with the parameters chosen above is designed to generate spectra with
 $r \approx 0.003$, over an order of magnitude below the current upper bound \cite{BICEP:2021xfz}. This value of $r$, together with the observed amplitude of temperature fluctuations in the CMB, requires that $V_0\approx 10^{-10} M_{\rm Pl}^4$ or, equivalently, $10^{-12}$ in non-reduced Planck units \cite{Ijjas:2013vea}. As power-law models have been eliminated by observations \cite{Ijjas:2013vea,balkenholInflationEnd20252025,Kallosh:2025sji}, plateau models have become the favored target of future observations.

There are two features of plateau potentials that appear to be problematic when \final{we take into account} the effects of anti-ultralocality. \final{Each of these features} can lead to \final{distinct} failure modes for plateau inflation:

\begin{itemize}
\item Compared to power-law models, there is a \final{significantly} larger \final{energy density} gap between
the big bang and the beginning of inflation in plateau inflation models. \final{This leads to} a much longer period of \final{spatial} gradient \final{growth} (anti-ultralocality) \cite{Ijjas:2013vea}. The growing spatial gradients \final{delay the onset of} inflation \final{and prevent} the universe from  achieving the 60 or more $e$-folds of homogeneous accelerated expansion \final{that are} required to explain the observations of the CMB.

\item The plateau region of the potential is extremely flat. As a result, the amplification of spatial gradients is nearly as likely to drive the field toward larger values of $\phi$ as it is to drive the field toward its potential minimum. In the `self-reproduction' regime where $\phi$ is large, quantum fluctuations dominate the evolution of the inflaton field. The result is quantum runaway \final{and the breakdown of the semi-classical approximation, which makes scientifically meaningful predictions impossible}.

\end{itemize}

In this paper, we use the tools of numerical relativity to study both failure modes.
We introduce an objective protocol involving two \final{distinct} tests to determine whether a given
model of inflation and set of initial conditions encounter either failure mode:
\begin{itemize}
\item {\it Initial State Test} \final{verifies} whether the initial conditions are \final{non-perturbatively far from} flat FRW, as expected when the universe first emerges from the big bang and the quantum gravity dominated phase.

\item {\it Successful Smoothing Test} \final{verifies} whether the accelerated expansion phase is sufficient\final{ly long} to smooth and flatten the universe to the \final{degree} needed to generate a \final{nearly scale-invariant} spectrum of scalar \final{density} perturbations consistent with observations \final{while avoiding} quantum runaway \final{and the breakdown of the semi-classical approximation that would make scientifically meaningful predictions impossible}.
\end{itemize}

 Our numerical studies show that models of plateau inflation appear to be unable to satisfy both tests simultaneously: Initial conditions that lead to successful inflation are \final{exponentially fine-tuned}, while simulations with generic initial conditions encounter one or both failure modes.

Our results are consistent with those reported in previous studies of plateau inflation~\cite{East:2015ggf, giannadakisCriticalValueInflationary2025,joanaCosmicInhomogeneitiesEarly2022,joanaInhomogeneousInitialConditions2021,Joana:2020rxm,eastherInitialConditionsSampling2013,brandenbergerInitialConditionsInflation2017,mullerInitialConditionsStarobinsky2025,mullerInitialConditionsStarobinsky2023,Elley:2024alx,Clough:2016ymm,Clough:2017efm,Corman:2022alv,launayStochasticInflationNumerical2025,Aurrekoetxea:2019fhr}. The key difference is that these earlier studies
did not consider whether their initial conditions, when extrapolated back \final{in time},
were \final{non-perturbatively far from} a flat FRW spacetime as expected on physical grounds, and/or whether the inflaton might
be driven into the self-reproduction regime. As a result, their conclusions about the success or
failure of inflation were more optimistic than those obtained here when applying the Initial State and Successful Smoothing Tests.

The paper is \final{organized} as follows. In Sec.~\ref{scheme} we describe the tetrad-based numerical relativity \final{scheme} used to generate the simulations in this study. \final{Our} code has been validated in previous investigations. Sec.~\ref{protocol} describes our test protocol.
Sec.~\ref{results} illustrates the two types of failure modes caused by anti-ultralocality. \final{We} demonstrate that they are encountered \final{when we begin with} generic initial conditions and quantify the fine-tuning of initial conditions required to avoid \final{failure}. In Sec.~\ref{discussion} we summarize our results and highlight a critical tension: reducing the problems created by anti-ultralocality requires increasing the inflationary energy scale, which is the opposite of what is needed to suppress the tensor-to-scalar ratio to the \final{degree} required by observations of the cosmic microwave background.

\section{Numerical relativity scheme}\label{scheme}

In this section, we briefly describe our numerical scheme for evolving and testing inflationary models, which has been validated in earlier studies; \final{see, \it{e.g.}, } \cite{Garfinkle:2023vzf,Ijjas:2020dws,Garfinkle:2008ei}.

 \final{Our numerical relativity simulation evolves} the (3+1) Einstein--scalar system of equations \cite{Buchman:2003sq} \final{in orthonormal-tetrad form}. As in prior works, we adopt a constant-mean-curvature time slicing:
\begin{align}
    \frac{d\ln |K|}{dt_{CMC}}\equiv -\frac{1}{\tilde{\mathcal{N}}_{max}(t_{CMC})},
\label{tCMC}
\end{align}
where $K/3$ is the mean extrinsic curvature and $\mathcal{N}$ is the lapse. $\tilde{\mathcal{N}}_{max}$ is the maximum value of $\mathcal{N}$ at time $t_{CMC}$ normalized so that $\frac{1}{3}|K| \tilde{\mathcal{N}}_{max}=1$. In the flat FRW limit, the time \final{coordinate} $t_{CMC}$ measures the maximum number of $e$-folds from the \final{onset} of inflation (denoted $N_e$ where applicable). For an FRW metric, $K= 3H$ where $H$ is the Hubble parameter.

We solve the Einstein--scalar equations in (3+1) dimensions but only present results for initial conditions with spatial variation along one dimension for computational efficiency. (In this and past studies, no significant discrepancy has been found between simulations with variations along 1 versus 2 spatial dimensions.)

The initial conditions were set using York’s conformal method \cite{York:1971hw} to ensure they satisfy the Hamiltonian and momentum constraints \cite{Garfinkle:2008ei,Ijjas:2021gkf}.
The spatial metric $\gamma_{ij}$ on the initial time slice ($t_{CMC}=t_0$) is \final{hence} conformally flat,
\begin{equation}
\gamma_{ij}(t_0,\vec{x}) = \psi^4(t_0,\vec{x})\delta_{ij},
\end{equation}
where $\psi$ denotes the conformal factor.
The two independent components of the spatial 3-curvature tensor ($\bar{n}_{ab}$ and $\bar{A}_b$) and the tetrad vector components (${\bar{E}{}_a}^i$) then satisfy:
\begin{alignat}{2}
\label{eq3}
&\bar{n}_{ab}(t_0,\vec{x}) &&= 0 , \\
&\bar{A}_b(t_0,\vec{x}) &&= -2\psi^{-1}{\bar{E}{}_b}^i\partial_i\psi,\label{eq4} \\
& {\bar{E}{}_a}^i(t_0,\vec{x}) &&= \psi^{-2}(K_0/3)^{-1}{\delta_a}^i,
\end{alignat}
where
$\bar{A}_b \equiv {\textstyle \frac12} \varepsilon_{b}{}^{cd}\bar{N}_{cd}$
is the antisymmetric part of $\bar{N}_{ab}$ (the nine intrinsic spatial curvature variables) and $\bar{n}_{ab} \equiv \bar{N}_{ab} - \varepsilon_{ab}{}^c \bar{A}_c$ is the symmetric part. An overbar denotes rescaling by the mean curvature ({\it i.e.}, dividing by \final{appropriate} powers of $|K|/3$ such that variables are dimensionless). Early alphabet indices ($a$, $b$, {\it etc.}) are frame indices and mid-alphabet indices ($i$, $j$, {\it etc.}) are coordinate indices.

 Deviations from \final{conformally flat} initial conditions are generated in subsequent evolution steps by the non-linear interaction terms in the Einstein equations. However, because our simulations require periodic boundary conditions, regions of positive curvature must be accompanied by regions with negative curvature and a region of negligible curvature in between. These constraints strongly favor inflation, since spatial curvature decays more slowly than other contributions (in the homogeneous limit) and can be most effective in interfering with the onset of inflation. As we will show, the advantage afforded by these constraints is \final{still} not enough to prevent anti-ultralocality effects from suppressing inflation.

There remains the freedom to specify the initial scalar field distribution $\phi(t_0,\vec{x})$, the conformally rescaled initial scalar field velocity
$\bar{Q}(t_0,\vec{x})\equiv\psi^{6}(t_0,\vec{x})\bar{\dot\phi}(t_0,\vec{x})$, and the divergence-free (transverse, traceless) part of the conformally rescaled shear tensor, $\bar{Z}_{a b}^{TT} (t_0,\vec{x})\equiv \psi^6(t_0,\vec{x})\bar{\Sigma}^0_{ab}(t_0,\vec{x})$. These quantities are parameterized as follows:
\begin{align}
\label{QQ2a}
\phi(t_0) &= f_1 \cos{\left({\textstyle \frac{m_1}{L}} x + d_1\right)} +
f_3 \cos{\left({\textstyle \frac{m_3}{L}} y + d_3\right)} +
\phi_0,\\
\label{QQ2b}
\bar{Q}(t_0) &=\left({\textstyle \frac{K_0}{3}}\right)^{-1}\times \Big(f_0 \cos{\left({\textstyle \frac{m_0}{L}} x + d_0\right)}
+ f_2 \cos{\left({\textstyle \frac{m_2}{L}} y + d_2\right)} + Q_0
 \Big)
\end{align}
where $K_0/3$ is the initial mean curvature, $L$ is the length of the periodic simulation box (in units of $(K_0/3)^{-1}$),
$Q_0, \phi_0, f_0, f_1,f_2, f_3, m_0, m_1, m_2, m_3, d_0, d_1, d_2, d_3$ are constants,
and
\begin{equation}
\tensor*{\bar{Z}}{_a_b^{TT}} = \left({ \textstyle \frac{K_0}{3}}\right)^{-1}
{
\renewcommand*{\arraystretch}{1.3}
\begin{pmatrix}
b_2 &
& \kappa &
& 0 \\
\kappa &{\;}
& a_1 \cos{\left({\textstyle \frac{m_z x}{L}}+\alpha_x\right)}+ b_1 &
& a_2 \cos{\left({\textstyle \frac{m_z x}{L}}+\alpha_x\right)} + c_3 \\
0 &
& a_2 \cos{\left({\textstyle \frac{m_z x}{L}}+\alpha_x\right)} + c_3 & {\;}
& - \tensor*{\bar{Z}}{^{TT}_1_1} - \tensor*{\bar{Z}}{^{TT}_2_2}
\end{pmatrix}},
\label{ZZ2}
\end{equation}
where $a_1, a_2, b_1, b_2, c_3, \alpha_x$ and $\kappa$ are constants. (The labeling convention in Eqs.~\ref{QQ2a}-\ref{ZZ2} matches the one given in \cite{Ijjas:2024oqn} for simulations in two dimensions but setting $a_3=c_1=c_2=0$ for the one dimensional case considered here.)

\section{Testing Protocol}\label{protocol}

\final{We introduce} a protocol of tests for evaluating whether a cosmological model can smooth and flatten the universe beginning from generic initial conditions \final{that are sufficiently} far from flat FRW. Our protocol \final{involves} two \final{tests}:
\begin{itemize}
\item \final{Initial State} Test: \final{verifies}
whether the initial conditions are generic;
\item \final{Successful Smoothing} Test: \final{verifies} whether the \final{resulting} accelerated expansion avoids quantum runaway and the multiverse, and whether it produces at least 60 $e$-folds of accelerated expansion after smoothing and flattening the universe, as required for quantum fluctuations of the inflaton to generate the spectral band of density fluctuations observed in the CMB.
 \end{itemize}

\subsection{\final{Initial State Test} }

\begin{enumerate}
    \item
 \textbf{All of} \{\boldmath$\max(\hat{C}_0)$, \boldmath$\max(\hat{P}_0)$,
    \boldmath$\bar\sigma_{\hat{C}}^{0}$, \boldmath$\bar\sigma_{\hat{P}}^{0} \} $ \textbf{are} $ \geq \mathcal{O}(1)$\unboldmath. This test \final{verifies} that the initial conditions are \final{non-perturbatively far from} flat FRW, consistent with the chaotic conditions expected after the universe emerges from the quantum gravity dominated phase following a big bang. We quantify this \final{condition} using the gauge/frame invariant measures known as the {\it Weyl scalar} ($C$) and the {\it Chern-Pontryagin invariant} ($P$) derived from the Weyl tensor, an approach introduced for this purpose in \final{Ref.~}\cite{Ijjas:2023bhh} and used in previous numerical relativity studies \cite{Ijjas:2023dnb,Garfinkle:2023vzf,Elley:2024alx,Ijjas:2024oqn}. The tests compare $C$ and $P$ to the mean curvature: $\hat{C} \equiv C/(\frac{1}{9} K^{4})$ and $\hat{P} = P/(\frac{1}{9} K^{4})$. A flat FRW spacetime has $\hat{C}=\hat{P}=0$; \final{spacetimes non-perturbatively far from} flat FRW corresponds to $\hat{C}$ and $\hat{P}$ greater than ${\cal O}(1)$ \cite{Ijjas:2023bhh}.
    
    Note that the maximum denotes the largest value reached within the simulation box, while the standard deviations $\sigma_C$ and $\sigma_P$ are computed across the entire box. The standard deviations are included to flag \final{special} cases where $\hat{C}_0$ and/or $\hat{P}_0$ are large but nearly uniform across the box.
    
    \item \boldmath$\lambda_i^0/(|K_0|/3)^{-1}<1$\unboldmath, where $\lambda^0_i$ is the wavelength of the spatially varying initial contribution $\Omega_i^0$, where $i \in[\dot\phi^2,(\nabla\phi)^2,\Sigma^2,k]$ labels the initial contributions from the scalar kinetic energy, scalar gradient energy, shear, and spatial curvature contributions to the mean curvature, respectively. This condition ensures these contributions vary significantly over the scale of the initial mean curvature $|K_0|/3$.
    \item \boldmath$(\delta\dot\phi/\bar{\dot{\phi}})\big|_{t_{\rm CMC}=0} \gtrsim 1$\unboldmath, where $\bar{\dot{\phi}}$ is the mean and $\delta\dot\phi$ is the root-mean-square variation of $\dot\phi$ across the simulation \final{box}. This condition ensures that the field velocity has no preferred direction up or down the plateau potential, as expected when the universe emerges from a quantum-gravity-dominated phase at an energy density much greater than the potential energy density of the plateau.
     
\end{enumerate}

\subsection{Successful Smoothing Test}
\begin{enumerate}
    \item {\bf The final 60 $e$-folds of inflation must have} \boldmath$\hat{C} <10^{-10}$ \textbf{and} $\hat{P}<10^{-10}$\unboldmath \textbf{ over at least one Hubble patch}. Inflation must last at least 60 $e$-folds during which the universe has smoothed and flattened to the extent that any remaining classical curvature variations are small compared to the mean square curvature fluctuation amplitude (${\cal O}(10^{-10})$) generated by quantum fluctuations of the inflaton.
    \item {\bf During accelerated expansion, Hubble patches with $\hat{C} \ll 1$ and $\hat{P} \ll 1$ must not enter the quantum-runaway regime} in which quantum fluctuations dominate over the classical evolution of the inflaton. Following Ref. \cite{Creminelli:2008es}, we define quantum runaway as having occurred if the condition
     \begin{align}
    B_{\text{multi}}\equiv \frac{3}{2\pi^2}\frac{H^4}{\dot\phi^2} >1
    \label{multib}
    \end{align}
    is sustained for at least an $e$-fold of inflation (i.e., a Hubble time). \final{Quantum runaway} is problematic for \final{the following} reasons. Although the plateau potential and its specific parameters were chosen to produce a particular outcome that matches cosmological observations, quantum runaway invalidates the semi-classical approximation from which that outcome is derived. \final{Taken at face value}, \final{quantum runaway leads to} an infinite range of outcomes \final{(dubbed the cosmological multiverse)} that are \final{each} different from the intended outcome, and are therefore inconsistent with cosmological observations. Because there is no probabilistically preferred outcome, a \final{cosmological} multiverse makes it impossible to determine empirically whether inflation ever occurred \cite{Vilenkin:1983xq,steinhardt_1983}.

\end{enumerate}

\section{Anti-ultralocality and plateau models \final{of inflation}}\label{results}

\begin{figure}[h]
    \centering
    \includegraphics[width=0.8\linewidth]{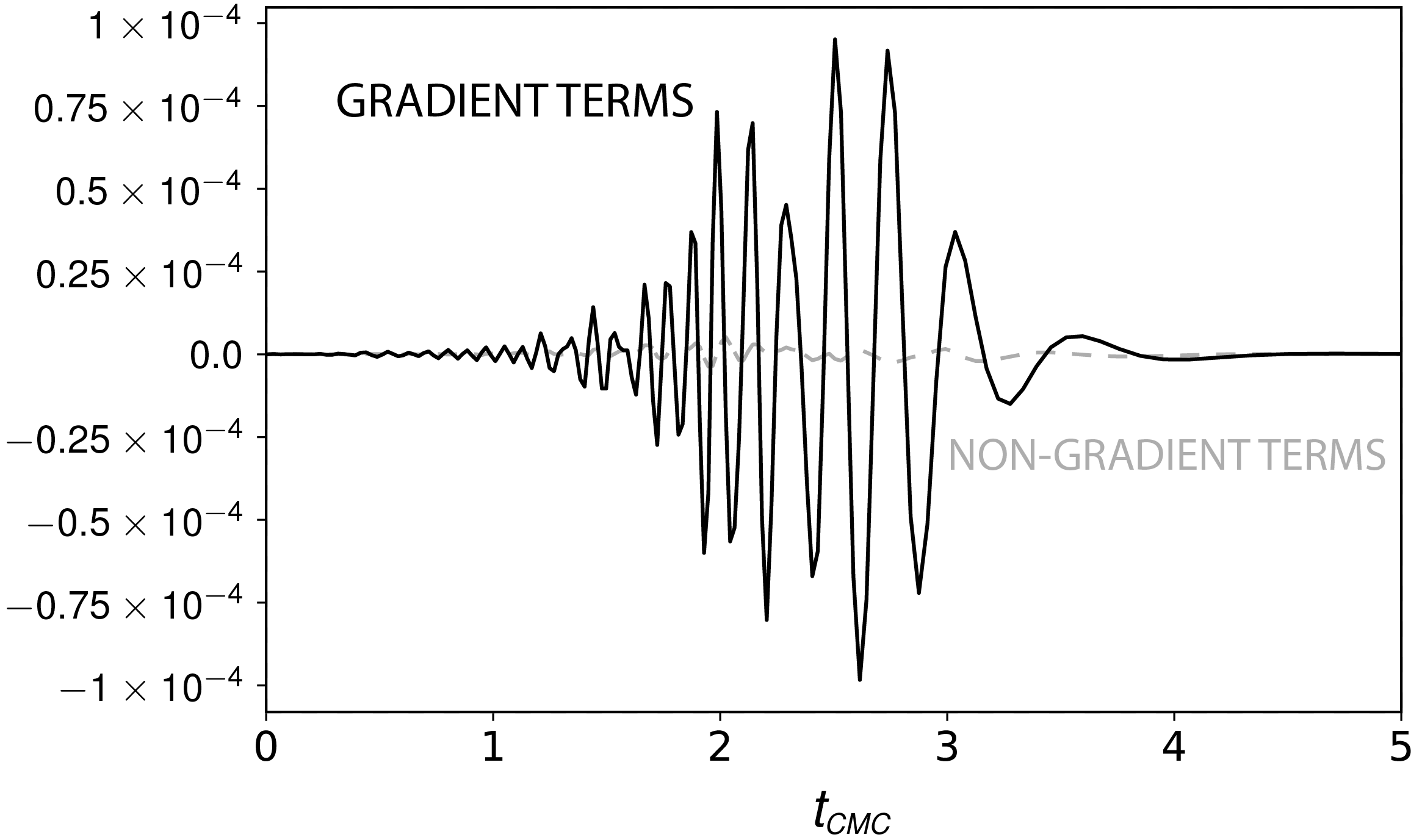}
    \caption{The evolution of gradient and \final{velocity} terms of $\tilde{\partial}_t \tensor{\bar{n}}{_1_1}$ for a representative spatial point for the example shown in Figure \ref{fig:3}. The sum of the gradient terms (solid black curve) grows to dominate the sum of the \final{velocity} terms (gray dashed curve) by an order of magnitude after just a few $e$-folds of expansion, indicating anti-ultralocality.
    }
    \label{fig:1}
\end{figure}

 After the universe emerges from the big bang, \final{the evolution becomes anti-ultralocal, {\it i.e.}, spatial gradients rapidly grow relative to velocities}. For example,
the equation of motion for the symmetric part of the intrinsic spatial curvature $n_{ab}$ is:
\begin{align}
    \tilde{\partial}_t \tensor{\bar{n}}{_a_b} = - \Big(\tilde{\cal N}-1\Big)\tensor{\bar{n}}{_a_b} +\tilde{\cal N}\Big( 2 \tensor{\bar{n}}{_(_a^c} \tensor{\bar{\Sigma}}{_b_)_c} - \uline{\tensor{\epsilon}{_(_a^c^d}\tensor{\bar{D}}{_c} \tensor{\bar{\Sigma}}{_b_)_d} }\Big)- \uline{\tensor{\epsilon}{^c^d_(_a}\tensor{\bar{\Sigma}}{_b_)_c}\tensor{\bar{D}}{_d}\tilde{\cal N} },
\end{align}
where $\Sigma_{ab}$ is the divergence-free part of the traceless shear tensor, $\epsilon_{abc}$ is the Levi-Civita symbol, $D_a$ is the directional derivative along the spatial orthonormal tetrad $e_a$, and subscript parentheses denote symmetrization (see the Appendix of \cite{Garfinkle:2023vzf}). The two underlined terms contain spatial gradients, while the remaining terms do not. \final{As illustrated} in Figure~\ref{fig:1}, after just a few $e$-folds of expansion, \final{the evolution turns anti-ultralocal:} the gradient terms grow more rapidly than the \final{velocity} terms and come to dominate (excepting moments when the former cross through zero).

\final{Note that in an expanding universe all physical quantities, including gradients and velocities, shrink with growing mean curvature, albeit at different rates. In our scheme,  all physical quantities are rescaled by appropriate powers of the mean curvature $K$. The rescaled, dimensionless variables may grow or shrink and they may do so at different rates. The terms that grow fastest correspond to physical quantities that shrink at the slowest rate as the universe expands and hence will come to dominate the total energy density. For example, if velocity terms grow fastest, the universe becomes ultralocal and homogeneous. But if gradient terms grow fastest, the universe becomes anti-ultralocal and inhomogeneous.}

In the following subsections, we show \final{examples} that \final{illustrate} anti-ultralocal \final{evolution during early decelerated expansion and discuss how it} leads to one of two critical
failure modes \final{for inflation}: either insufficient smoothing and flattening \final{or self-reproduction}.
 \final{These examples are representative of the many} simulations we ran with \final{a wide range of} initial conditions.

\subsection{Failure mode 1: gradients prevent sufficient smooth\final{ing} }\label{41}

 \final{We evolve the Einstein-scalar system using} the following parameters for the initial \final{field distribution and velocity configuration defined in Eqs.~(\ref{QQ2a}-\ref{QQ2b}):} 
\begin{alignat}{4}
    f_1&=0.05,&\quad m_1&=8,&\quad d_1&=-1.05,&\quad\phi_0&=7,\label{param10}\\
    f_0&=3\times 10^{-2},&\quad m_0&=2,&\quad d_0&=-1.57,&\quad Q_0&=-5\times 10^{-2}.
\end{alignat}
\final{For the shear variables in Eq.~\eqref{ZZ2}, we choose}
\begin{alignat}{4}
    a_1&= 3.3\times 10^{-3},&\quad a_2&= 1.0\times 10^{-3},&\quad b_1&= 1.88\times 10^{-3},&\quad b_2&=-1.5\times 10^{-4},\\
    c_3&=-1.4\times 10^{-4},&\quad \kappa&=10^{-6},&\quad m_z&=3,&\quad \alpha_x&=\alpha_y=0
    \label{param20}.
\end{alignat}
\final{For the height of the potential plateau, we choose} $V_0=10^{-4} (K_0^2/3)$, and \final{our simulation box length} is $L=10(K_0/3)^{-1}$.

\begin{figure}[h]
    \centering
    \includegraphics[width=0.7\linewidth]{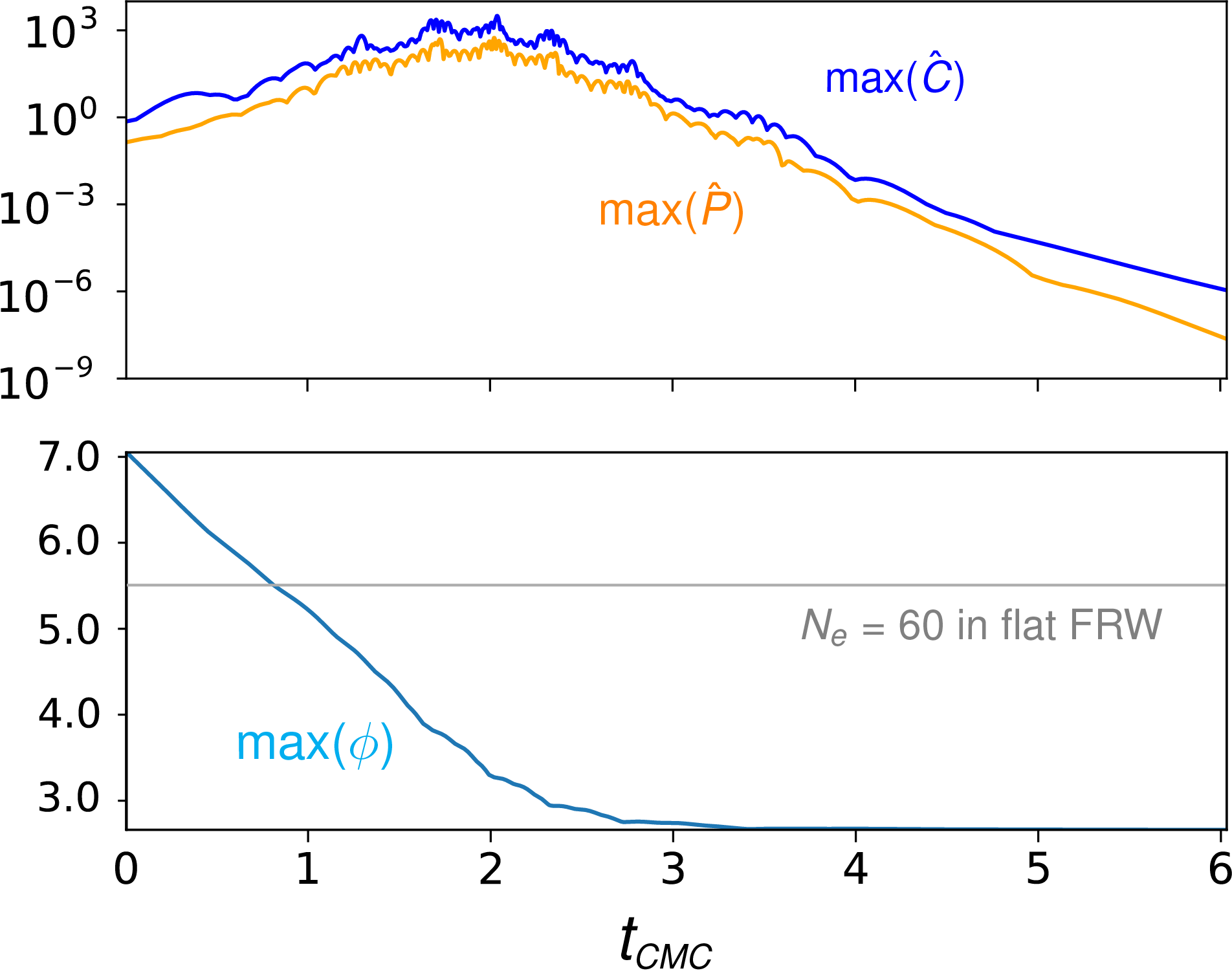}
    \caption{Maximum values of $\hat{C}$ and $\hat{P}$ (top) and $\phi$ (bottom) \final{across the entire simulation} box for a plateau height of $V_0 = 10^{-4} (K_0^2/3)$ and parameters specified in Eqs.~(\ref{param10}-\ref{param20}). $\hat{C}$ and $\hat{P}$ grow sharply due to anti-ultralocality, during which time $\phi$ rolls significantly down its potential without driving any accelerated expansion. The horizontal line at $\phi = 5.45M_{\rm Pl}$ denotes the minimum field value required for 60 $e$-folds of nearly-flat FRW inflation, but by the time $\hat C$ and $\hat P$ drop to $\mathcal{O}(10^{-10})$, $\phi$ has fallen well below this line and \final{the remaining inflation is not sufficient} to satisfy the 60 $e$-fold requirement.}
    \label{fig:2}
\end{figure}

The maximal values of $\hat C$ and $\hat P$ as a function of $t_{CMC}$ are shown in the upper panel of Figure~\ref{fig:2}. Their growth by several orders of magnitude at the beginning of the simulation \final{demonstrates that the evolution has turned} anti-ultralocal during the phase of decelerated expansion preceding inflation.
This increases the time needed for inflation to suppress $\hat C$ and $\hat P$ below $\mathcal{O}(10^{-10})$, until which point the initial classical inhomogeneities are too large and the deviation from flat FRW is too great to generate a nearly scale-invariant spectrum of quantum fluctuations of the inflaton.

If less than 60 $e$-folds of inflation remain when $\hat C$ and $\hat P$ fall below $\mathcal{O}(10^{-10})$, then it is not possible to generate a nearly scale-invariant spectrum of density perturbations in accord with CMB observations.
 For the plateau model in Eq.~(\ref{potential}), sufficient $e$-folds remain only if
 $\phi > 5.45 M_{\rm Pl}$ when $\hat C$ and $\hat P$ fall below $\mathcal{O}(10^{-10})$. As the bottom panel in Figure~\ref{fig:2} shows, $\phi$ fails to reach this mark and the plateau model fails this test.

Note that in this example, we have set $V_0=10^{-4} (K_0^2/3)$ in order to reduce the hierarchy between the Planck scale and the inflationary energy scale for numerical tractability. In plateau models \final{that are compatible with observations}, \textit{i.e.,} with $V_0=10^{-10} (K_0^2/3)$, the phase of anti-ultralocal expansion lasts even longer, making such models even more susceptible to this failure mode.

\pagebreak
\subsection{Failure mode 2: \final{The growth of} gradients trigger\final{s} quantum runaway and a multiverse \final{of outcomes}}\label{42}

To avoid the first failure mode studied in Sec.~\ref{41}, one may think to increase \final{the initial average field value $\phi_0$} and/or reduce the initial magnitude of \final{the average field velocity $Q_0$}, so inflation may last longer and achieve the $60$ $e$-fold requirement while still satisfying \final{the conditions of the Initial State} Test. In this subsection, however, we show that a second failure mode is then encountered: the combination of anti-ultralocality and the flatness of the plateau potential results in \final{a breakdown of the semi-classical approximation, which leads, in turn, to} eternal self-reproduction and a multiverse \final{of outcomes that violate the second condition of the Successful Smoothing Test}.

Our numerical code is purely classical, so the quantum fluctuations responsible for self-reproduction are not explicitly included. Rather, as we describe\final{d} in Sec.~\ref{protocol}, we track the evolution of $B_{\rm multi}$ \final{defined above} in Eq.~(\ref{multib}) during the phase of accelerated expansion to determine whether it exceeds unity across a region of the box at least as large as a Hubble radius for at least a Hubble time.

 \final{We consider an example where the initial scalar field velocity and spatial distribution satisfy the Initial State Test} and  the resulting evolution achieves 60 or more  $e$-folds of inflation, avoiding the failure mode of Sec.~\ref{41}:
\begin{alignat}{4}
    f_1&=0.12,&\quad m_1&=13,&\quad d_1&=-1.05,&\quad\phi_0&=10,\label{param1}\\
    f_0&=3\times 10^{-4},&\quad m_0&=17,&\quad d_0&=-1.57,&\quad Q_0&=-5\times 10^{-4}.
\end{alignat}
\final{For the shear variables in Eq.~\eqref{ZZ2}, we choose}
\begin{alignat}{4}
    a_1&= 3.3\times 10^{-4},&\quad a_2&= 1.0\times 10^{-4},&\quad b_1&= 1.88\times 10^{-4},&\quad b_2&=-1.5\times 10^{-5},\\
    c_3&=-1.4\times 10^{-5},&\quad \kappa&=10^{-7},&\quad m_z&=14,&\quad \alpha_x&=\alpha_y=0.
    \label{param2}
\end{alignat}
The \final{simulation} box size is $L=10(K_0/3)^{-1}$. By setting initial conditions such that the field is initially rolling down the plateau at all points within the simulation, this example \textit{favors inflation} because the field starts out moving away from the large-$\phi$ self-reproduction regime.

\begin{figure}[h]
    \centering
    \includegraphics[width=0.7\linewidth]{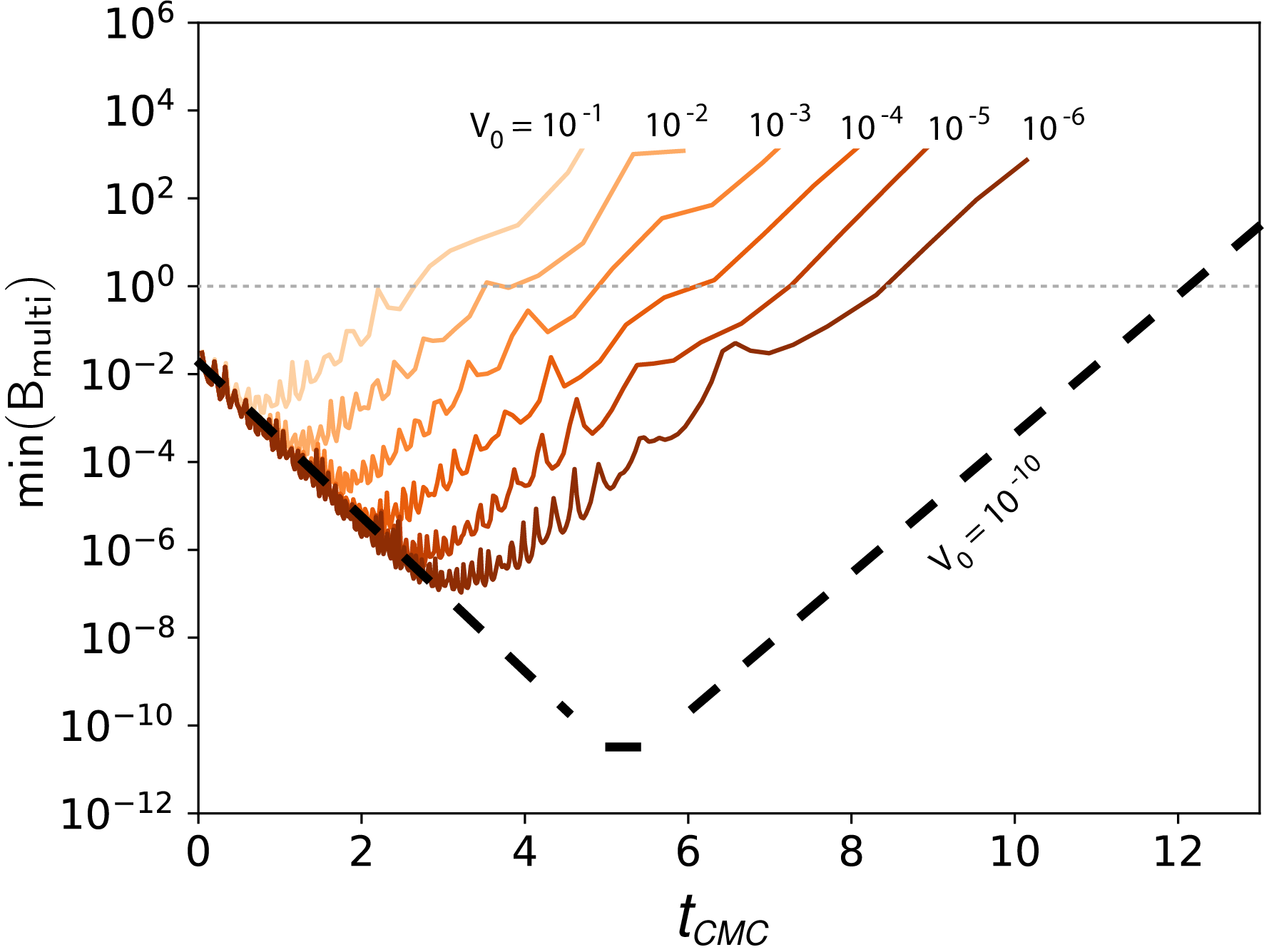}
    \caption{The evolution of the minimum value of $B_{\rm multi}$ across the simulation for plateau heights $V_0$ ranging from $10^{-1}-10^{-6}(K_0^2/3)$ and the parameters in Eqs. ~(\ref{param1} - \ref{param2}). The data (solid lines) are extrapolated to the target value of $V_0=10^{-10}(K_0^2/3)$ (black dashed line). When $\min(B_{\rm multi})>1$ for at least a Hubble time ($ \Delta t_{CMC}\sim 1$, gray dotted line), the entire simulated volume enters the self-reproduction (multiverse) regime.
}
    \label{fig:3}
\end{figure}

During the phase of decelerated expansion when the potential is negligible, the evolution of $B_\mathrm{multi}$ is governed by the competition between the decreasing curvature scale $K/3 \sim H$ and the decreasing field velocity $\dot{\phi}$. (In this phase of expansion, $\phi(x)$ is effectively a free field whose dynamics obey a wave equation damped by a generalized Hubble friction, causing $\dot{\phi}(x)$ to oscillate with a decreasing amplitude over time.) While these terms individually decrease at comparable rates, $B_{\rm multi}\propto H^4/\dot\phi^2$ is affected by four powers of $H$ and hence initially decreases over time, as shown in Figure \ref{fig:3}.

Once the potential dominates and the phase of accelerated expansion begins, however, $H$ approaches a positive constant while $\dot\phi$ continues to decrease at roughly the same rate. This causes $B_{\rm multi}$ to increase, ultimately exceeding unity everywhere in the simulation box and triggering quantum runaway. Figure \ref{fig:3} shows that even as the scale of the potential is lowered to the target value of $V_0=10^{-10}(K_0^2/3)$,
self-reproduction is still reached well before 60 $e$-folds of inflation have been completed.

\subsection{Only initial conditions close to flat FRW can avoid both failure modes}\label{43}

Having demonstrated \final{how plateau models of inflation are prone to two distinct} failure modes, we now ask what initial conditions are required to fully satisfy \final{the Successful Smoothing Test, {\it i.e.},} \final{to obtain an early} phase of inflation \final{that} lasts at least 60 $e$-folds, gener\final{ates} the correct CMB temperature fluctuations, and never enters \final{the self-reproduction regime}. We find that only extraordinarily smooth initial conditions, with exponentially small values of both $\hat{C}_0  \lesssim 10^{-8}$ and $\hat{P}_0 \lesssim 10^{-14}$, can avoid these failure modes.
In terms of our protocol, this corresponds to strongly violating the criteria specified in \final{our Initial State Test} in order to satisfy the criteria of the \final{Successful Smoothing Test}.

\begin{figure}[h]
    \centering
    \includegraphics[width=0.7\linewidth]{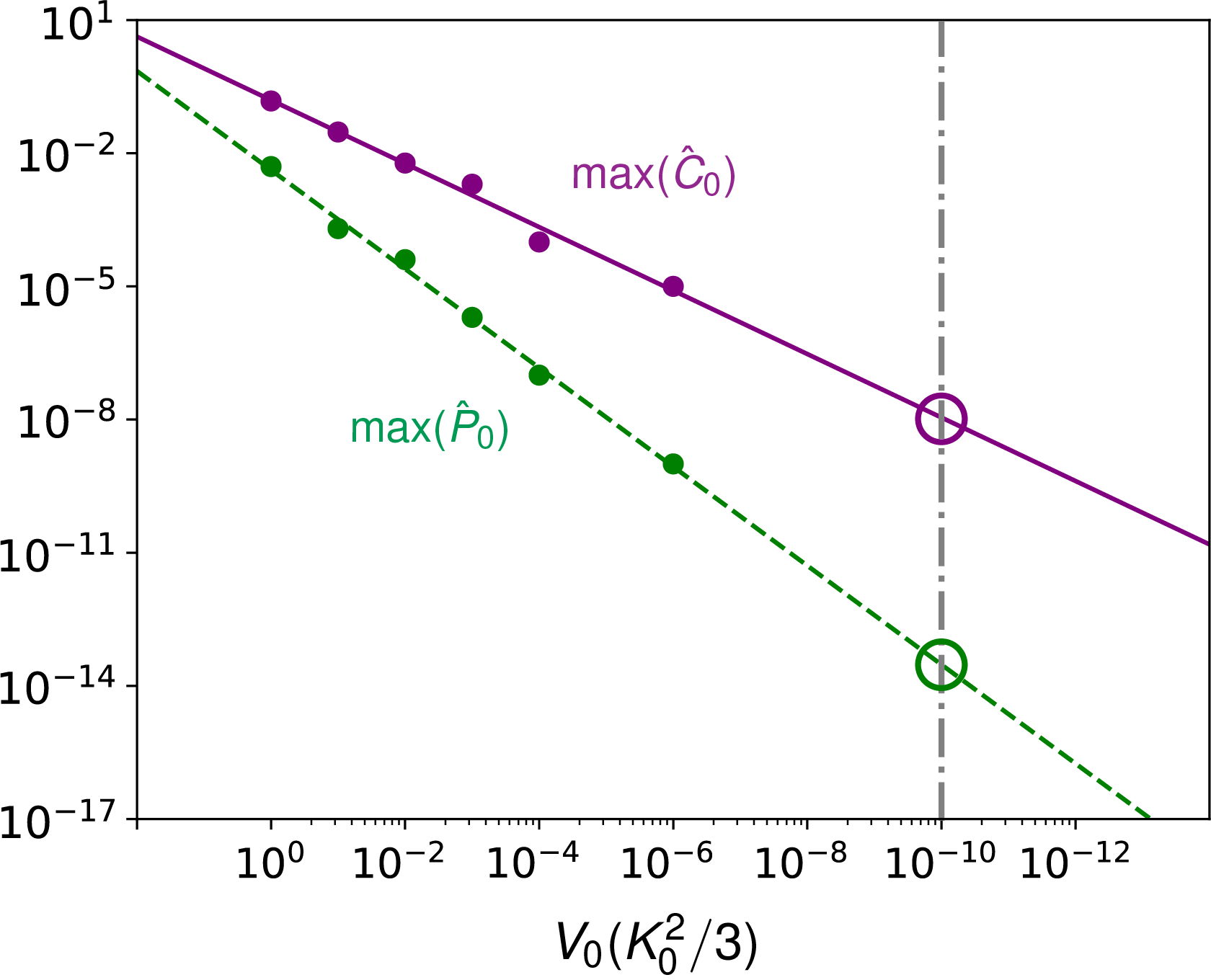}
    \caption{Required fine-tuning of $\hat{C}_0$ and $\hat{P}_0$ at the moment the universe emerges from the big bang, as a function of the plateau height $V_0$. Solid circles represent simulation results and open circles represent the extrapolated values for the case $V_0 = 10^{-10} (K_0^2/3)$, the value required to obtain an observationally acceptable tensor-to-scalar ratio. Note that $\hat{C}_0$ is less than $10^{-8}$ and $\hat{P}_0$ is less than $10^{-14}$ in this case.}
    \label{fig:5}
\end{figure}

\final{W}e have systematically studied how $\hat C_0$ and $\hat P_0$ must be progressively fine-tuned as the plateau potential height is decreased from the Planck \final{density} down to the level needed to obtain a tensor-to-scalar ratio consistent with observations, $V_0 = 10^{-10} (K_0^2/3)$. As $V_0$ decreases, the duration of decelerated expansion \final{before the onset of inflation} increases, extending the
period of anti-ultralocal \final{evolution} shown in Fig. \ref{fig:2}. To compensate, progressively smaller values of $\hat{C}_0$ and $\hat{P}_0$ are required to satisfy \final{the Successful Smoothing Test}, as \final{illustrated} in Fig.~\ref{fig:5}. Extrapolating this trend, we project that $\hat{C}_0$ and $\hat{P}_0$ must be less than $10^{-8}$ and $10^{-14}$ respectively for the target value of $V_0 = 10^{-10} (K_0^2/3)$ (shown as open circles in Fig.~\ref{fig:5}).

As a numerically tractable example with a potential height of $V_0=10^{-4}(K_0^2/3)$, the following parameters yield the greatest $\hat C_0$ and $\hat P_0$ out of all sampled initial conditions:
\begin{alignat}{4}
    f_1&=0.07,&\quad m_1&=30,&\quad d_1&=-1.05,&\quad\phi_0&=14,\label{param01}\\
    f_0&=3\times 10^{-5},&\quad m_0&=17,&\quad d_0&=-1.57,&\quad Q_0&=-10^{-3},\\
    a_1&=10^{-5},&\quad a_2&= 6\times 10^{-5},&\quad b_1&=1.8\times10^{-5},&\quad b_2&=-0.9\times 10^{-5},\\
    c_3&=-0.8\times 10^{-5},&\quad\kappa &=10^{-7},&\quad m_z&=19,&\quad \alpha_x&=\alpha_y=0.\label{param02}
\end{alignat}\noindent
This example uses a \final{simulation} box size \final{given by} $L=35(K_0/3)^{-1}$. Even at this higher plateau height, Fig. \ref{fig:5} shows that $\hat C_0$ and $\hat P_0$ must be tuned to at least the order of $10^{-4}$ to satisfy the Successful Smoothing Test.

The degree of initial smoothness and flatness required to obtain a tensor-to-scalar ratio consistent with CMB observations ($\hat C_0<10^{-8}$ and $\hat P_0<10^{-14}$ for $V_0=10^{-10}(K_0^2/3)$) is far from what is expected when the universe emerges from the big bang and \final{quantum gravity dominated} phase. In fact, it is comparable to the homogeneity, isotropy and spatial flatness that {\it were supposed to be the result of} inflation, not the prerequisite for it. \final{O}ur results suggest that for inflation to succeed, some other mechanism altogether may be required to solve the homogeneity, isotropy, and flatness problems of standard big bang cosmology.

\section{Discussion}\label{discussion}

The conventional view is that evolution in a successful inflationary cosmology proceeds as follows: 
\begin{enumerate}
\item \final{The initial geometry of the universe at the Planck scale, following the big bang, is non-perturbatively far from
a flat FRW spacetime, including significant inhomogeneities, anisotropies, and spatial curvature.}
\item \final{Starting from such conditions,} the inflaton field \final{evolves by the laws of classical general relativity. Its potential energy density comes to dominate the total energy density, sourcing} accelerated expansion that drives the universe toward a flat-FRW state and \final{for} at least $\sim 60$ $e$-folds thereafter.

\item During the final $\sim 60$ $e$-folds, quantum fluctuations of the field seed a nearly scale-invariant spectrum of primordial \final{curvature} fluctuations. \final{At the same time, quantum effects remain subdominant everywhere}, \final{such that the semi-classical approximation remains valid throughout and there is no self-reproduction}.
\end{enumerate}

In our numerical study of the plateau model in Eq.~(\ref{potential}), however, we have found no examples in which all three of these statements hold true. Using a pair of quantitative test\final{s} \final{that we introduced} in Sec.~\ref{protocol}, we have illustrated that generic initial conditions lead to either insufficient classical expansion (Sec.~\ref{41}) or quantum runaway (Sec.~\ref{42}). Furthermore, our results suggest that the desired \final{semi-classical} evolution of the inflaton field can only be achieved if the initial conditions at the Planck scale are already homogeneous and isotropic to a\final{n exponentially} high degree across causally disconnected regions (Sec.~\ref{43}).

We expect these results to generalize to any \final{sufficiently flat} potential that \final{has a} low \final{characteristic} energy scale \final{as} required by observational constraints on the tensor-to-scalar ratio \cite{BICEP:2021xfz,Lyth:1996im}. A corollary is that the scale of inflation cannot be reduced arbitrarily, as \final{is sometimes claimed,} because anti-ultralocality effects get significantly worse and require increasingly non-generic fine-tuned initial conditions.

\acknowledgments

\vspace{-0.5em}
\noindent
We would like to thank F. Pretorius for extensive discussions and valuable guidance. We would additionally like to thank N. Patino, C. Millett, S. Birmingham, C. Welsh, and K. Clough for helpful conversations. This work is supported in part by the U.S. Department of Energy under grant number DE-FG02-91ER40671 and by the Simons Foundation under grant number 654561. 


\appendix
\section{Convergence tests and other technical details}

In our simulations, we use a grid of 1024 points and a Courant factor of $0.5$. Our choice of gauge yields an elliptic equation for the lapse function. Our formulation of the Einstein-scalar system contains \final{a set of} hyperbolic equations for the evolution of the gravitational and scalar field variables, along with a set of corresponding constraint equations. \final{Our code evolves this hyperbolic-elliptic system.}
 \final{We use the} constraints \final{to verify the convergence of our code.} See \cite{Garfinkle:2023vzf} for more details and \cite{Ijjas:2020dws} for a helpful derivation of a historical version of our formulation. 

\begin{figure}[h]
    \centering
    \includegraphics[width=0.8\linewidth]{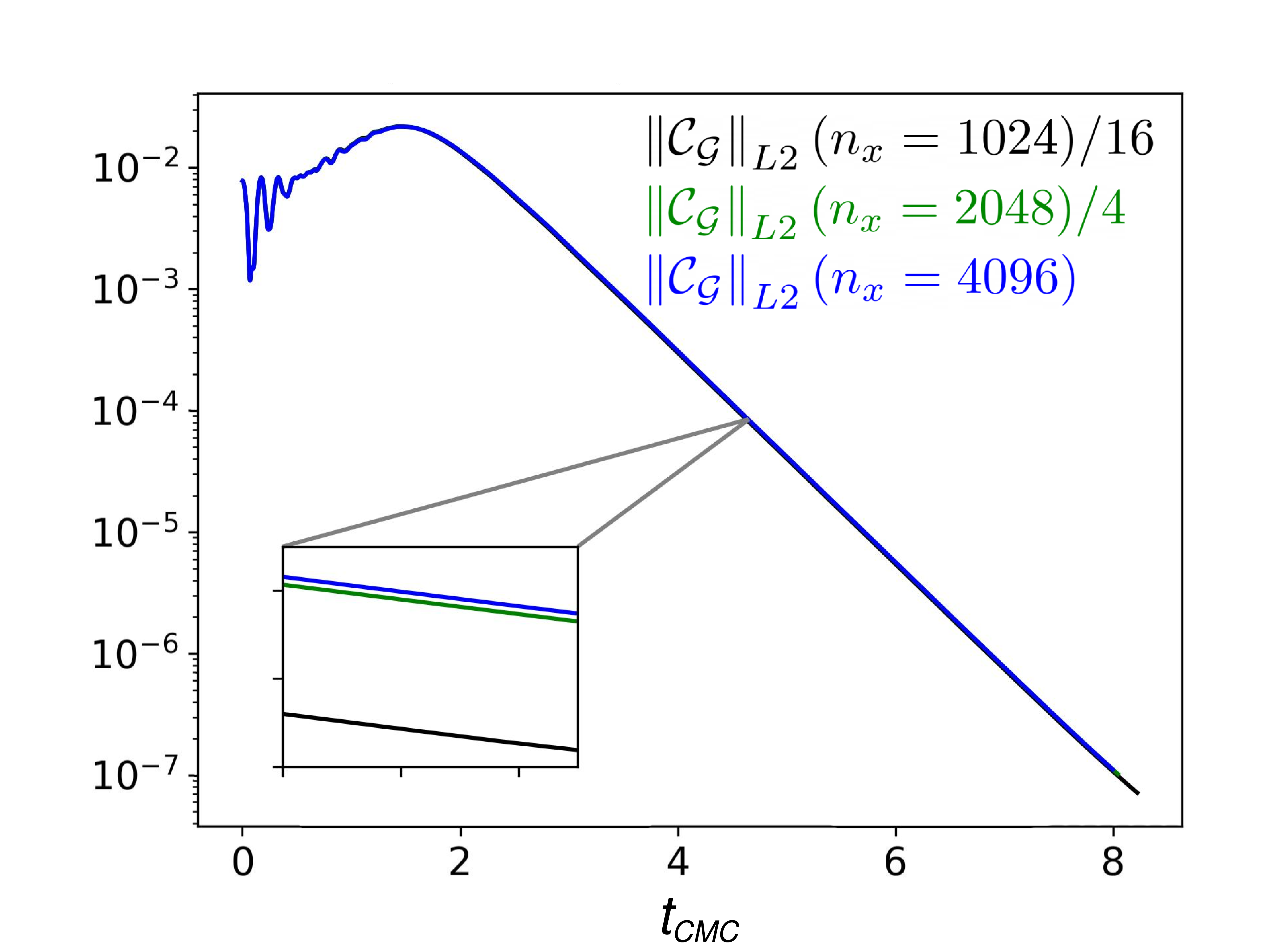}
    \caption{\final{A successful convergence test comparing the integrated Hamiltonian constraint as a function of time for three resolutions for the simulation shown in Figure \ref{fig:3} with $V_0=10^{-4}(K_0^2/3)$, where each curve is rescaled according to the resolution.}}
    \label{1_6}
\end{figure}

All \final{simulations} presented in this work satisfy all of the appropriate convergence tests for the Einstein-scalar system. As an example, Figure \ref{1_6} illustrates a convergence test for the evolution of the 2nd-order $L2$ norm of the Hamiltonian constraint integrated over the entire spatial domain of the test in Sec.~\ref{42}.

\phantomsection
\addcontentsline{toc}{section}{References}
\bibliographystyle{utphys}
\bibliography{references.bib}

\end{document}